\documentclass[aps,prl,superscriptaddress,twocolumn,longbibliography,floatfix]{revtex4-2}
\usepackage{units}
\usepackage{amsmath}
\usepackage{amsthm}
\usepackage{amssymb}
\usepackage{graphicx}
\usepackage{float}  
\usepackage[section]{placeins}
\usepackage[english]{babel}
\usepackage[utf8]{inputenc}
\usepackage{mathtools}
\usepackage{color}
\usepackage{xcolor}
\usepackage{bbold}
\usepackage{braket}
\usepackage{tabularray}
\usepackage{subfigure}

\usepackage[version=3]{mhchem}
\usepackage{dsfont}
\usepackage{ulem}
\makeatletter
\def\uwave{\bgroup \markoverwith{\lower3.5\p@\hbox{\sixly \textcolor{red}{\char58}}}\ULon}
\font\sixly=lasy6 
\makeatother
\definecolor{myurlcolor}{rgb}{0,0,0.7}
\definecolor{myrefcolor}{rgb}{0.8,0,0}

\usepackage[
	colorlinks=true, 
	linkcolor=myrefcolor,
	citecolor=myurlcolor,
	urlcolor=myurlcolor
]{hyperref}
\usepackage{cleveref}

\usepackage[linesnumbered, ruled,vlined]{algorithm2e}

\usepackage{ulem}
\makeatletter
\def\uwave{\bgroup \markoverwith{\lower3.5\p@\hbox{\sixly \textcolor{red}{\char58}}}\ULon}
\font\sixly=lasy6 
\makeatother

\graphicspath{{./images/},{./imagesAppendix/},{./Plots/}}

\renewcommand{\eqref}[1]{Eq.~(\ref{#1})}

\ifx\proof\undefined
\newenvironment{proof}[1][\protect\proofname]{\par
	\normalfont\topsep6\p@\@plus6\p@\relax
	\trivlist
	\itemindent\parindent
	\item[\hskip\labelsep\scshape #1]\ignorespaces
}{
	\endtrivlist\@endpefalse
}
\providecommand{\proofname}{Proof}
\fi

\makeatother

\providecommand{\ftname}{ft}
\providecommand{\theoremname}{Theorem}
\providecommand{\claimname}{Claim}
\providecommand{\lemmaname}{Lemma}
\providecommand{\definitionname}{Definition}

\definecolor{KB}{rgb}{0.4,0.3,0.9}

\definecolor{THc}{rgb}{0.9,0.3,0.2}

\newcommand{\sectionMain}[1]{
\let\oldaddcontentsline\addcontentsline
\renewcommand{\addcontentsline}[3]{}
\section{#1}
\let\addcontentsline\oldaddcontentsline
}

\allowdisplaybreaks

\begin{document}

\title{ 
Solitary waves of attracting SU($N$) fermions\\
}
\author{Aysha Alsubaihi}
\affiliation{Quantum Research Center, Technology Innovation Institute, P.O. Box 9639 Abu Dhabi, UAE}
\author{Wayne J. Chetcuti}
\affiliation{Universit\'e Grenoble Alpes, CNRS, LPMMC, 38000 Grenoble, France}
\author{Frederic Chevy}
\affiliation{Laboratoire de Physique de l’Ecole Normale Sup\'erieure, ENS, Universit\'e PSL,
CNRS, Sorbonne Universit\'e, Universit\'e Paris Cit\'e, F-75005 Paris, France}
\affiliation{Institut Universitaire de France (IUF), 75005 Paris, France}
\author{Juan Polo}
\affiliation{Quantum Research Center, Technology Innovation Institute, P.O. Box 9639 Abu Dhabi, UAE}
\author{Luigi Amico}
\affiliation{Quantum Research Center, Technology Innovation Institute, P.O. Box 9639 Abu Dhabi, UAE}
\affiliation{Dipartimento di Fisica e Astronomia ``Ettore Majorana'', Universit\'a di Catania, Via S. Sofia 64, 95123 Catania, Italy}
\affiliation{INFN-Sezione di Catania, Via S. Sofia 64, 95123 Catania, Italy}

\begin{abstract}
    We study the formation, dynamics, and disorder robustness of bound states in attractively interacting SU($N$) fermions on a one-dimensional ring lattice. Using exact diagonalization in fixed-momentum sectors and Bethe ansatz exact results as a guide, we resolve the many-body spectrum into bands related to the possible partitions of the particles into bound composites, and characterize their internal structure also through density-density and $N$-body correlations. A pinning quench protocol reveals a transition from dispersive spreading to dynamical localization as the attractive interaction increases relative to the single-particle hopping. We find that the bound state dynamics for one-particle per component occurs as a many-body quantum walk similar to that of a single particle with a re-normalized effective mass. Such a property, that is the quantum version of the shape-preserving motion of classical solitons, can provide the dynamical signature of the fermionic solitary waves. We probe the robustness of the fermionic bound-state dynamics under on-site disorder.
\end{abstract}

\maketitle

\textit{Introduction --} 
Correlations in quantum many-particle systems are  resources for quantum simulation, sensing, and computation~\cite{Ac_n_2018, Koch2025}. Particularly relevant for achieving quantum advantage are those quantum states characterized by correlations with no classical counterpart~\cite{lamprou2025recent,aasi2013enhanced,li2018quantum,kleiner2004superconducting}. 
In this context, self-bound localized solitary quantum waves, propagating without spreading in space, play a distinguished role~\cite{kivshar1989dynamics,mazets2006different,Naldesi_2019}. Ultracold bosonic gases with  attractive atom-atom interactions, realized experimentally for example by $^7$Li or $^{85}$Rb ~\cite{Khaykovich_2002, Strecker_2002,marchant2013controlled}, provide an important route to realize solitary quantum matter-waves in the form of bright solitons.
Robustness against dispersion, together with the ability to create their specific superpositions, make quantum solitary waves natural building blocks for matter-wave interferometry and atomtronic devices~\cite{cuevas2013_063006,weiss2009creation,PhysRevLett.113.013002,Naldesi2022, Polo2021, Polo2020,lee2006enhanced,amico2022colloquium,polo2024perspective}.

Extending the paradigm of solitary waves to fermions is inherently subtle, as the Pauli exclusion principle naturally opposes the high-density localization required for many-particle self-trapping~\cite{yefsah2013heavy} (see also \cite{coleman1975quantum}). On the other hand,  the manipulation of this feature could open a path to the rich variety of quantum phases of interacting fermions that are inaccessible in purely bosonic systems.
To this end, a feasible avenue  is provided by multi-component degenerate Fermi gases. Indeed, by utilizing different atomic hyperfine states serving  as `spin' degree of freedom, the fundamental restriction of Pauli blocking is weakened.
Multi-component fermions with flavor-symmetric interactions, known as SU($N$) fermions, can be realized in ultracold atom experiments using alkaline-earth and ytterbium isotopes, where the nuclear spin provides the flavor (`spin') degree of freedom~\cite{Sonderhouse_2020, gorshkov2010two, Song2020, Scazza_2014}. SU($N$) fermions trapped in optical lattices have opened an unprecedented frontier in quantum simulation, realizing Hubbard-like models~\cite{gorshkov2010two}; in  the latter, the  multi-components  generically increase the degeneracy of the states, thus  suppressing  classical thermal fluctuations while magnifying quantum fluctuations~\cite{cazalilla2009ultracold,toth2010three,corboz2011simultaneous,hermele2009mott,Capponi_2016,Cazalilla_2014,ibarra2025many}. 

In the attractive regime,  the SU($N$) Fermi-Hubbard model~\cite{Capponi_2016, Cazalilla_2014}, supports  a hierarchy of bound states. These multi-component bound states span  from conventional Cooper-like pairs to complex composite clusters of  $N$ distinct flavors~\cite{rapp2007color}. These exotic states provide an ideal macroscopic platform to probe non-perturbative pairings, phase separation boundaries, and color-flavor locking analogs of  high-energy dense quark matter~\cite{baym2010bcs}.
In the dilute lattice regime, such bound states can be rigorously classified through  the  Gaudin-Yang-Sutherland integrable model~\cite{Sutherland1968, Takahashi1970,Chetcuti_2023}. Recent work has characterized these states through correlation functions and through their signatures in persistent currents on ring geometries~\cite{polo2025persistent,Polo2020,Naldesi2022,Polo2021,pecci2021probing,chetcuti2023probe,chetcuti2023interference,polo2026static}.

The goal of the present work is to track the features of bright soliton matter-waves in a mesoscopic degenerate gas of $N_p$ attracting SU($N$) fermions in a one-dimensional ring lattice. We compute the exact many-body energy spectrum via exact diagonalization in fixed-momentum sectors and use exact Bethe ansatz results in the  dilute regime to gain a deeper understanding of the composite states. 
The spectrum of the system results to be organized into energy bands corresponding to  different bound-state decompositions separated from scattering states  characterized  by the presence of unbound particles. These decompositions may involve composites  such as pairs, trions, quartets or more generally $N$-body bound states. We shall see that, in contrast to bosons, the lowest fermionic sub-band exhibits an additional fine structure imposed by the Pauli exclusion principle, which becomes progressively less restrictive as the number of components increases, causing the spectrum to approach the bosonic limit. The internal structure of bound states and energy bands of the system is tracked through exact means in the dilute limit of the lattice system, and corroborated through density-density and $N$-body correlations in the general case. Then we will demonstrate how the above features reflect in the quench dynamics of the system after a localized state is pinned~\cite{Naldesi_2019,blain2026quantum}. In particular, we show that many-body bound states can indeed propagate as quantum walks, a feature that has recently been established for bosonic bright solitons in the quantum realm \cite{blain2026quantum}.  Finally, we study the bound state dynamics under on-site disorder.

\textit{Models and Methods --} To model $N_p$ attractively interacting $N$-component fermions trapped in an $L$-site ring-shaped lattice, we employ the SU($N$) Fermi-Hubbard Hamiltonian
\begin{equation}
\hat{H} = -J \sum_{i=1}^{L} \sum_{\alpha}^{N} \left( \hat{c}^\dagger_{i,\alpha} \hat{c}_{i+1,\alpha} + \text{h.c.} \right) + U \sum_{i=1}^{L} \sum_{\alpha < \beta}^{N} \hat{n}_{i,\alpha} \hat{n}_{i,\beta}
\label{model}
\end{equation}
where $\hat{c}_{i, \alpha}^{\dagger}$ ($\hat{c}_{i, \alpha}$) creates (annihilates) a fermion of flavor $\alpha \in \{1, \dots, N\}$ on site $i$, and $\hat{n}_{i, \alpha} = \hat{c}_{i, \alpha}^{\dagger}\hat{c}_{i, \alpha}$. The parameters $J$ and $U$ denote the hopping amplitude and on-site interaction strength respectively. We consider attractive interactions $U < 0$, fix $J$ as our energy scale, and impose periodic boundary conditions $\hat{c}_{L+1, \alpha} \equiv \hat{c}_{1, \alpha}$ throughout. Our analysis covers $N \in \{2, 3, 4\}$, which are accessible in alkaline earth-like and ytterbium atom experiments~\cite{Sonderhouse_2020, Scazza_2014,pagano2014one}.\par

In the continuous limit of vanishing lattice spacing, or, equivalently, the dilute lattice regime $\nu = N_p/L \ll 1$, the physics of the system (\ref{model}) is captured by the Gaudin-Yang-Sutherland integrable  model describing SU($N$) fermions mutually interacting through  the potential $c \sum_{i<j}\sum_{\alpha,\beta}\delta (x_{i,\alpha}-x_{j,\beta})$, $c$ being the effective attraction~\cite{Sutherland1968, Takahashi1970,Chetcuti_2023}.
In this limit, the exact energies can  be accessed exactly through Bethe ansatz in terms of the so-called rapidities $k$s and $\Lambda$s through a set of coupled (Bethe) equations.  
Away from the dilute limit, the  model (\ref{model})  for $N>2$ is not Bethe ansatz integrable. 
The way we characterize the bound states is two-fold. 
First, we exploit the translational symmetry of $\hat{H}$ and block-diagonalize it in sectors of fixed  lattice momentum $k_l = 2\pi l/L$, with $l = 0, 1, \dots, L-1$. Collected across momentum sectors, the spectrum organizes into dispersive bands $E(k)$. A band whose energy lies below the scattering continuum and separated from it by a finite gap is identified as a bound-state band. The size of the gap measures how strongly the state is bound (binding energy), and provides a measure of their stability. We therefore use the appearance of isolated bands, and the growth of their separation with increasing $|U|/J$, as the spectral signature of bound-state formation.

To characterize these states and resolve the internal structure of each band, we compute the density-density correlations and $N$-body correlations. The first are defined as
$G(r) = \langle \hat{n}_{0,\alpha} \hat{n}_{r,\beta} \rangle,$ for $\alpha \neq \beta $
where $G(r)$ probes the spatial coincidence of distinct flavors at the pairwise level. For an $m$-flavor bound state, all flavors cluster at the same site, producing a sharp peak in $G(r)$ at $r = 0$ that decays rapidly with distance. A scattering state of unbound fermions, by contrast, gives a nearly uniform $G(r)$ consistent with uncorrelated occupation. The $N$-body correlator we study is 
$T_N(r) = \Big\langle \hat{n}_{j_0+r,\,1}\,\prod_{\alpha=2}^{N} \hat{n}_{j_0,\alpha} \Big\rangle,$ which fixes $N-1$ flavors at the reference site $j_0$ and measures the probability that the remaining flavor is found at distance $r$. At $r=0$ this reduces to the on-site $N$-body density $T_N(0) = \langle \prod_{\alpha=1}^{N} \hat{n}_{j_0,\alpha} \rangle$, which is non-zero only when all $N$ flavors occupy the same site. Together, these correlators provide a direct fingerprint of the bound-state composition in each band.

To assess whether the bound states behave as self-trapped objects, we perform a quench protocol. The system is prepared in the ground state of $\hat{H} + \hat{H}_\mathrm{pin}$, where
$\hat{H}_\mathrm{pin} = V_\mathrm{pin}\sum_{\alpha}\hat{n}_{j_{0}, \alpha}$ applies an attractive potential, $V_\mathrm{pin}<0$, at a central site $j_0$. The pinning strength $V_\mathrm{pin}$ is chosen to be of the order of the bandwidth of the lowest bound band, sufficient to localize the particles near $j_0$ without significantly distorting the internal bound-state structure. At time $t = 0$ the pinning is removed and the state evolves under $\hat{H}$ in (\ref{model}). We track the local density $n_j(t) = \sum_\alpha 
\langle \hat{n}_{j,\alpha}(t) \rangle$. The robustness of the bound state to disorder is probed by adding an on-site potential $\hat{H}_\mathrm{dis} = \sum_{i,\alpha} \epsilon_i \hat{n}_{i,\alpha}$ where the local energies $\epsilon_i$ are drawn independently from a uniform distribution on $[-W, W]$ at each disorder realization, where $W$ is the maximum strength of the disorder.

\begin{figure}[h]
    \centering
    \includegraphics[width=\columnwidth]{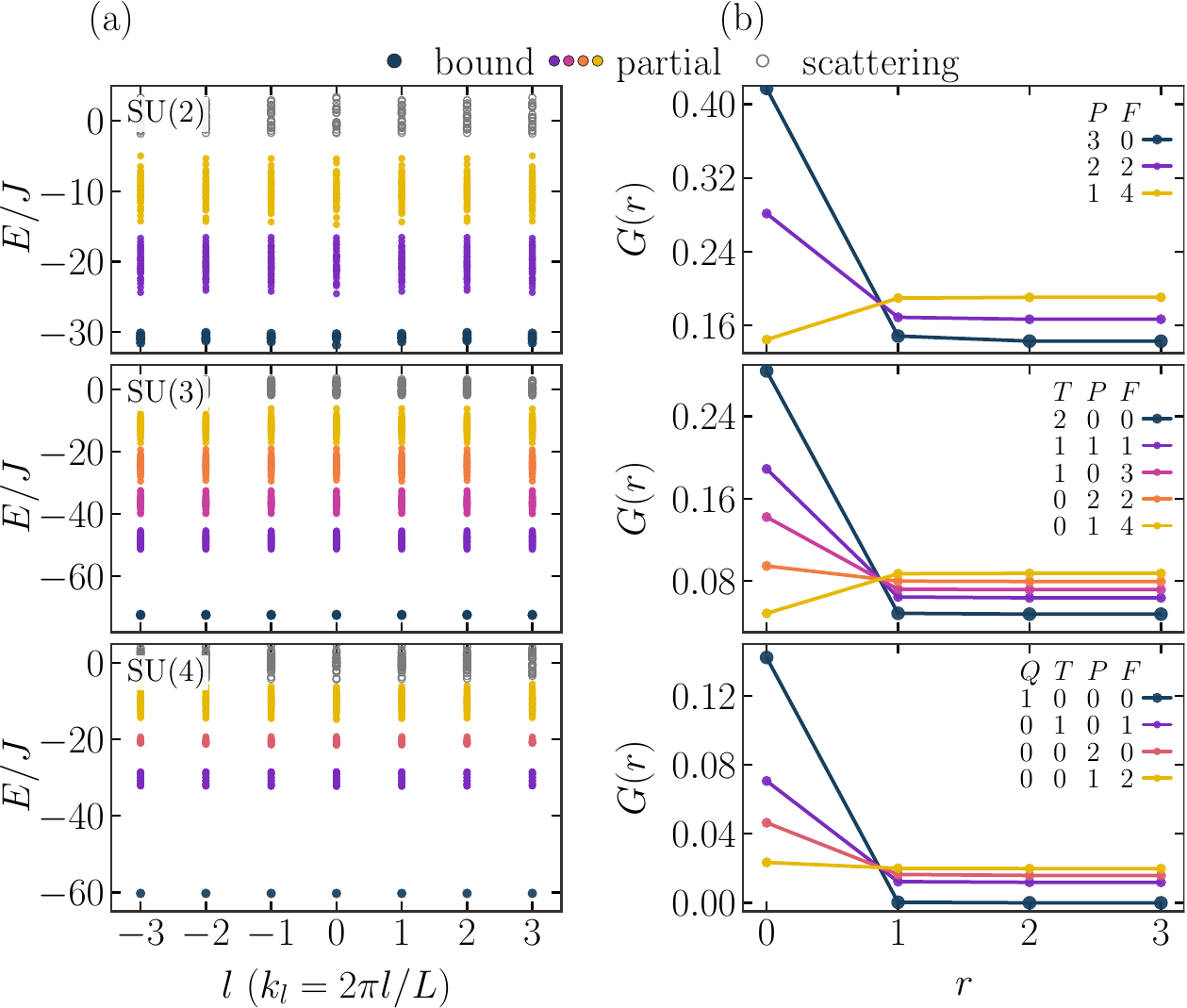}
    \caption{Energy-momentum spectrum $E(k_l)$ (left column) and cross-flavor density-density correlator $G(r)$ (right column) for SU(2) $N_p = 6$ at $|U|/J = 10$, SU(3) with $N_p = 6$ at $|U|/J = 12$, and SU(4) with $N_p = 4$ at $|U|/J = 10$, on $L = 7$ sites. Colors encode the binding hierarchy: navy denotes the maximally bound lowest-energy band, the gradient denotes progressively more dissociated partially bound configurations, and gray markers denote scattering states. Each $G(r)$ curve is averaged over all eigenstates of the corresponding band, within the zero-momentum ($k_l=0$) sector. The right-column legends list the composite content, where $P$, $T$, $Q$, and $F$ denote pairs, trions, quartets, and unpaired fermions, respectively.}
    \label{fig:fig1}
\end{figure}

\textit{Spectrum and static correlations --} Figure~\ref{fig:fig1} shows the many-body energy spectra and cross-flavor density-density correlation function $G(r)$ for SU(2), SU(3), and SU(4) fermions on a ring with $L = 7$ sites. We consider SU(2) with $N_p=6$ at $|U|/J =10$, SU(3) with $N_p=6$ at $|U|/J=12$, and SU(4) with $N_p=4$ at $|U|/J=10$. The spectral gaps as functions of the interaction strength are presented in the Appendix. As shown in Fig.~\ref{fig:fig1}(a), the spectra are organized into bands separated by interaction-induced gaps. Each band is related to a different partition of the particles into bound composites and unbound fermions. The lowest bands contain the maximally bound configurations: three pairs for SU(2), two trions for SU(3), and one quartet for SU(4). At higher energies, these composites progressively dissociate into smaller clusters and unbound fermions. The highest bands, shown in grey, represent the fully unbound scattering states.

Important insight into the structure of the sub-bands and the nature of the bound states can be gained by analyzing the dilute regime of the system. In this limit, the spectrum can be  found exactly through the Bethe ansatz of the Gaudin-Yang-Sutherland model. The ground state is a bound state characterized by the so-called  $k-\Lambda$ string solutions  (of the Bethe equations). 

Each string solution describes a bound state composed of $m$ fermions, $m \in \{1,2,\ldots,N\}$~\cite{Chetcuti_2023,guan2013fermi}. For fixed $k$ configurations, parameters $\Lambda$s provide states with different energies   (corresponding to holon-type excitations in the `spin' sector).   
The energies of a state composed of various bound states of size $m$ reads
$ 
E=\sum\nolimits_{m=1}^{N}\sum\nolimits_{j=1}^{N_{m}}(mp_{j,m}^{2} - I_{m})
$ 
where $p$ is a suitable function of $k$'s and $\Lambda$'s and $I_{m}$ is given by
$ 
    I_{m} = {m(m^2-1)}c^{2}/3
$ - see the Appendix~\ref{eq:string}. 
The arguments above show that indeed, sub-bands arise  by the degree of the particles' clustering parametrized by the possible excitations in the `spin' sector: 
The lowest energy sub-band of bound states  is characterized by a bound state of $N$ particles, while the higher energies  bands are combinations of `lower order' obtained binding progressively $N-1, N-2, \dots, 1$ particles. 

We extrapolate the structure above to  the generic, non-integrable case Eq. (\ref{model}). The bands corresponding to lower order bound states are found to be separated by gaps that grow with $|U|$, reflecting the increasing energetic cost of breaking higher-degree composites into smaller clusters. 

Indeed, the number of bands is determined by the distinct allowed ways of partitioning $N_{p}$ particles into $m$-composites, subject to the statistical constraint that each composite contains at most one fermion of each flavor.  We also note that for $N_p/N=1$, the lowest band contains $L$ states, while for larger occupation $N_p/N>1$ the number of states in the lowest band increases.

The density-density correlator $G(r)$ confirms the assignment of bands to bound-state configurations, as shown in Fig.~\ref{fig:fig1}(b). Each curve is obtained by averaging $G(r)$ over all eigenstates assigned to a given band within the zero-momentum $k_l=0$ sector in the lattice. The fully bound configurations exhibit the largest on-site correlations and the most rapid decay with distance. As the composites dissociate into unbound fermions, $G(r)$ decreases and the correlations become broader. The fully unbound scattering states, shown in grey in Fig.~\ref{fig:fig1}(a), have nearly uniform correlations consistent with uncorrelated occupation and are omitted from Fig.~\ref{fig:fig1}(b) for clarity. A more detailed eigenstate-resolved analysis of the density-density correlation function $G(r)$ across the momentum sectors is presented in the Appendix.

\textit{Dynamics --} We investigate the time evolution of these bound states when they are initially prepared in a spatially localized region of the lattice. Specifically, the system is initialized in the ground state of $\hat{H} + \hat{H}_\mathrm{pin}$, which localizes the particles around a central site $j_0$. The pinning is then removed, and the state evolves under the unrestricted Hamiltonian $\hat{H}$ Eq.~\ref{model}. For sufficiently strong attractive interactions and short time scales, a self-trapped composite remains localized near $j_0$ throughout the evolution, while for weaker interactions, where the states are no longer protected by a gap, the density distribution shows how the particles spread across the lattice. We find that the interaction is magnified by  $N$, affecting the localization properties of the bound states.

\begin{figure}[h]
    \centering
    \includegraphics[width=\columnwidth]{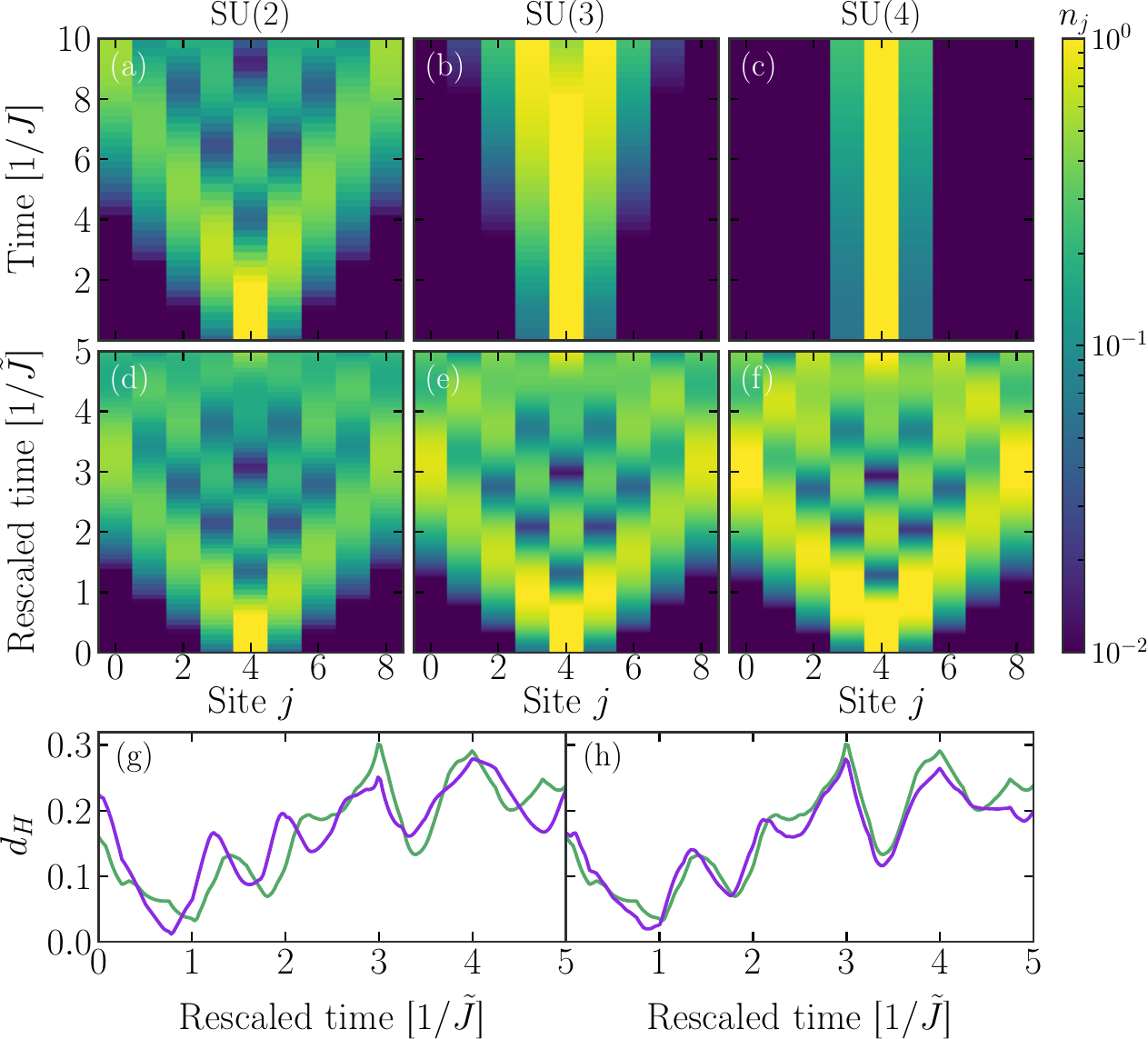}
    \caption{Density evolution $n_j(t)=\sum_\alpha \langle \hat n_{j,\alpha}(t)\rangle$ following the release of a local pinning potential in the attractive $\mathrm{SU}(N)$ Fermi--Hubbard ring with $L=9$ and one particle per component $N_p/N=1$. Columns correspond to $(N,N_p)=(2,2)$, $(3,3)$, and $(4,4)$. The product $|U|N/J=12$ is fixed. Panels \textbf{(a)--(c)} show the density evolution in time $t$ (units of $1/J$) and panels \textbf{(d)--(f)} show the evolution as a function of the dimensionless rescaled time $\tau=t\tilde J$. \textbf{(g--h)} Hellinger distance $d_H$ between the composite density and the corresponding single-particle quantum walk, versus rescaled time, for SU($2$) (green) and SU($3$) (purple). Panel \textbf{(g)} compares the two systems at fixed $|U|/J$, whereas panel \textbf{(h)} compares them at matched $|U|N/J=12$. The small values of $d_H$ show that the bound composite approximately follows a single-particle quantum walk after rescaling by $\tilde J$.}
    \label{fig:fig2}
\end{figure}

The evolution is shown in Fig.~\ref{fig:fig2}(a--c), in the absence of disorder. For SU(2) [panel (a)], the density released from $j_0=(L-1)/2$ spreads across the ring within a few hopping times. For SU(3) [panel (b)], the density spreads more slowly and a clear maximum remains at $j_0$ while part of it moves outwards. For SU(4) [panel (c)], the density remains concentrated in a narrow region around $j_0$ throughout the time window shown, portraying the localization of the bound state. The degree of spreading decreases as $N$ increases. 

We note however that for $N_p/N=1$, the dynamics can be rescaled by a characteristic effective hopping $\tilde J \simeq J\frac{N}{(N-1)!}\,  \left(J/|U|\right)^{N-1}$~\cite{mansikkamaki2022beyond},
which reduces to the familiar pair hopping $\tilde J = 2J^2/|U|$ for $N=2$. $\tilde J$ provides us with a timescale made explicit in Fig.~\ref{fig:fig2}(d--f), where time is measured in units of $1/\tilde J$. This effective hopping rescales the dynamics to an effective composite particle with an increased effective mass.

In the rescaled time $\tau = t\tilde J$, the density evolutions shown in [Fig.~\ref{fig:fig2}(d--f)] become nearly identical: particles spread from $j_0$ into a checkerboard profile, corresponding to the interference provided by a quantum walk~\cite{Preiss2015,cai2021multiparticle}. The $N$-dependence is accounted for by the scaling of $\tilde J$, and once it is rescaled away, the composites evolve in the same manner, that is the dispersive evolution of a single particle with a rescaled effective mass.   

To quantify how closely the bound composite reproduces this single-particle quantum walk, we compute the Hellinger distance~\cite{PhysRevA.97.062342} between the composite density and that of a free particle [Fig.~\ref{fig:fig2}(g--h)]. At each rescaled time, denoting the normalized densities as $p_j$ (composite) and $q_j$ (single particle), the Hellinger distance is $d_H = \frac{1}{\sqrt2}\sqrt{\sum_j\left(\sqrt{p_j}-\sqrt{q_j}\right)^2}$. We find that the Hellinger distance remains smaller for the bound-state dynamics than for generic many-body dynamics involving unbound particles during almost the entire evolution - see Appendix.

\begin{figure}[h]
    \centering
    \includegraphics[width=\columnwidth]{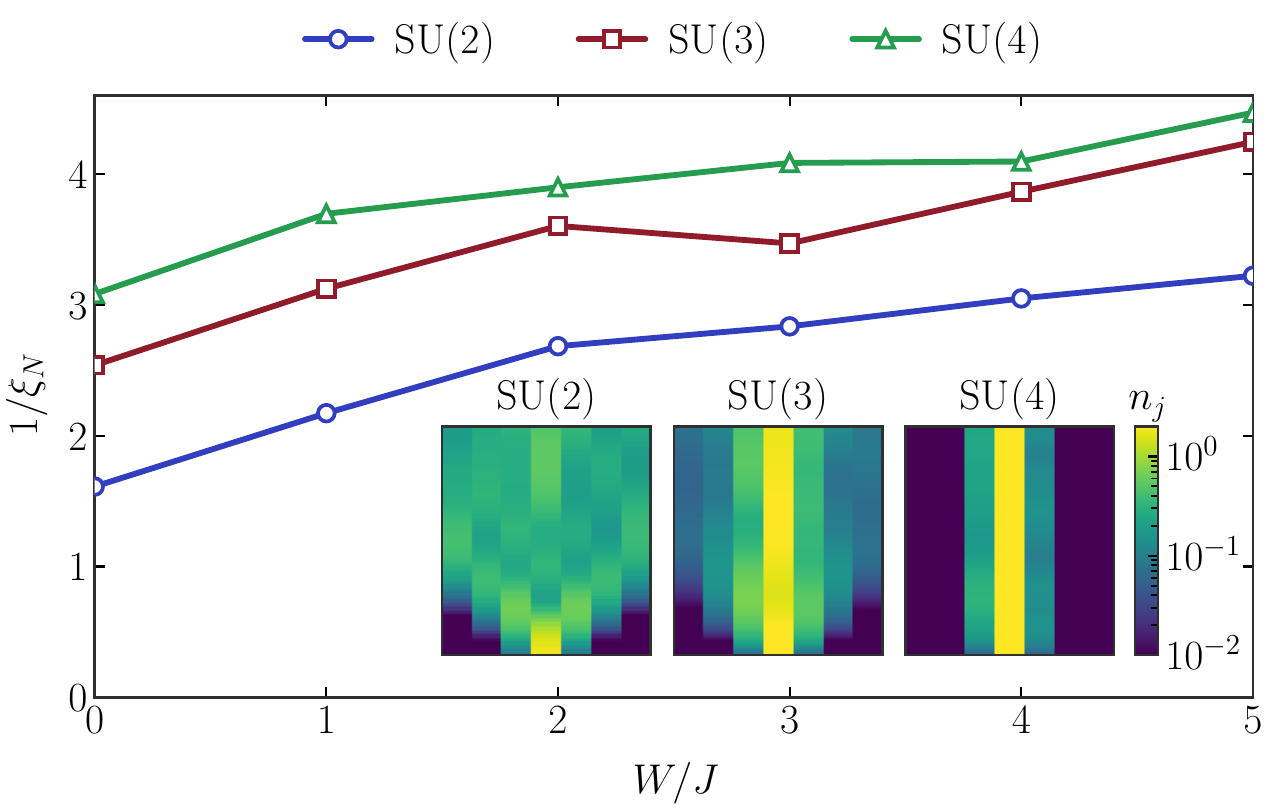}
    \caption{Robustness against on-site disorder at matched $|U|N/J=12$ and one particle per component $N_p/N = 1$. Main panel: inverse correlation length $1/\xi_N$ of the $N$-body correlator $T_N(r)$ vs. disorder strength $W/J$, for SU(2), SU(3), and SU(4) on an $L=7$ ring. For each $W/J$, the on-site point $r=0$ is excluded and $1/\xi_N$ is obtained from the magnitude of a linear fit to $\log T_N(r)$ for $r\geq1$. The inverse correlation length displays an overall increasing trend with disorder and remains larger for higher $N$. Insets: representative disorder-averaged density evolution $n_j(t)$ following the pinning quench at $W/J=0.25$ on an $L=7$ ring, shown with a logarithmic color scale in the rescaled time $t\tilde J$ for SU(2), SU(3), and SU(4).}
    \label{fig:fig3}
\end{figure}

\textit{Disorder --}
We now probe the robustness of the bound-state dynamics against on-site disorder by adding $\hat{H}_{\mathrm{dis}}$. We work at one particle per component ($N_p/N=1$) and average all quantities over independent disorder
realizations. We first follow the density under weak disorder [insets of Fig.~\ref{fig:fig3}], $W/J=0.25$, in rescaled time. The quantum walk is suppressed unequally over the window shown. The SU(2) density continues to spread, while the SU(3) and SU(4) densities stay concentrated near $j_0$. The insets show the disorder-averaged density dynamics following the pinning quench, whereas the main panel reports the inverse correlation lengths extracted from the disorder-averaged ground-state $N$-body correlator.

This behavior is quantified by the ground-state $N$-body correlator
$T_N(r)$. For each value of $W/J$, we first average $T_N(r)$ over the
disorder realizations and then fit $\log T_N(r)$ linearly for $r\geq1$,
excluding the on-site peak at $r=0$. The magnitude of the fitted slope
defines the inverse correlation length,
$1/\xi_N=|\mathrm{d}\log T_N/\mathrm{d}r|$, shown in
Fig.~\ref{fig:fig3}. The inverse correlation length displays an overall
increasing trend with disorder for all three SU($N$) and remains larger
for higher $N$. Thus, the $N$-body correlations become more tightly
confined in relative distance under disorder, with a stronger response
for larger $N$. Together with the density evolution in the insets, this
indicates stronger localization of the composite dynamics.

\textit{Conclusion --} 
We conclude that, indeed, the balancing act between statistical exclusion in SU($N$) fermions and interspecies attraction can give rise to bound states acting as  localized multi-component solitary wave. By studying the dynamics of the  bound states, we find that they move indeed as a quantum walks, this feature providing the quantum analog of the shape-preserving dynamics of classical solitons.  

To validate this statement, we analyzed the system's spectral properties and dynamics.

The Hamiltonian many-body spectrum is found to be organized in specific sub-bands characterized by particle clustering and their internal correlations (see Fig.~\ref{fig:fig1}). In particular, for the generic case of  $N>N_p$,  the  physics of the bound state is governed by two energy scales which are the distance to the scattering states and the width of the bound states sub-bands. The latter is found to scale inversely with particles' interactions and $N$. 
The organization of the sub-bands is established by combining exact spectral results  (Bethe ansatz of the Gaudin-Yang-Sutherland) and numerical calculation of correlation functions.

A quench protocol uncovers a  crossover from dispersive motion to dynamical localization, in which increasing $N$ leads to markedly enhanced localization - see first row of Fig~\ref{fig:fig2}. A specific scaling is found, reflected in the time dynamics - see middle row of Fig~\ref{fig:fig2}. By studying the Hellinger distance between single and many-body propagation, we demonstrate that the bound state dynamics defines indeed a quantum walk: bound solitary waves evolve according to  single particle type of  particle-wave duality~\cite{Preiss2015}. Such feature should be read as the quantum analog of the shape-preserving motion of the classical solitary wave.
The effect of disorder is studied - see Fig~\ref{fig:fig3}.
Strikingly, by studying the $N$-point correlations, the bound state dynamics  is found robust against on-site disorder. This feature is quantified through  correlations that results to stay sharp and even steepen with $W$ (Fig.~\ref{fig:fig3}).

Fermionic solitary waves' intrinsic stability against dispersion and disorder we established here suggests that such bound composites can act as protected carriers for quantum technology.  In particular,  in interferometric setups in which beam splitters can be implemented as suitable local potential offsets along the guide~\cite{chih2021reinforcement,Polo2021,naldesi2023massive,polo2026static, Polo_2013}, fermionic solitons can provide a  resource: their composite nature and internal correlations enable multi-particle phase coherence that could enhance sensitivity beyond single-particle interference, while their resistance to decoherence mechanisms improves signal stability in realistic environments. 

\let\oldaddcontentsline\addcontentsline

\renewcommand{\addcontentsline}[3]{}

\medskip

\textit{Acknowledgements --} We thank G. Marchegiani, G. Catelani, A. Minguzzi, and B. Blain for discussions. WJC received funding from the European Union’s Horizon research and innovation programme under the Marie Sk\l odowska-Curie grant agreement \textit{SUN\textunderscore Atomtronics} (no. 101205763).

\bibliography{references} 
\clearpage
\onecolumngrid
\raggedbottom
\appendix
\section{Structure of the bound states}
\subsection{String solutions of the Gaudin-Yang-Sutherland model}

Following the continuum-limit mapping presented in Ref.~\cite{Chetcuti_2023}, the Gaudin--Yang--Sutherland model is obtained from the SU($N$) Hubbard
model by taking the continuum limit in which the lattice spacing
vanishes.

Let $D$ denote the particle density on a lattice of $L$ sites with
spacing $\Delta$, such that
$D=N_p/(L\Delta)$. The filling factor can therefore be written as
$\nu=N_p/L=[N_p/(L\Delta)]\Delta$. Hence, when
$\Delta\rightarrow 0$ at fixed particle density $N_p/(L\Delta)$, the
filling factor also approaches zero. To preserve the fermionic anticommutation relations in the continuum
limit, the lattice operators are rescaled according to
$c_{i,\alpha}^{\dagger}
=\sqrt{\Delta}\Psi_{\alpha}^{\dagger}(x_i)$ and
$n_{i,\alpha}
=\Delta\Psi_{\alpha}^{\dagger}(x_i)\Psi_{\alpha}(x_i)$,
where $x_i=i\Delta$. The continuum field operators $\Psi_{\alpha}$ and
$\Psi_{\alpha}^{\dagger}$ satisfy the standard anticommutation relations
$\Psi_{\alpha}(x),\Psi_{\beta}^{\dagger}(y)\}
=\delta_{\alpha,\beta}\delta(x-y)$ and 
$\{\Psi_{\alpha}^{\dagger}(x),\Psi_{\beta}^{\dagger}(y)\}=0.$

Through a Taylor expansion and integration by
parts~\cite{amico2004universality}, the SU($N$) Hubbard Hamiltonian is
mapped onto the continuum Fermi-gas theory:
\begin{equation}
\label{eq:hh3}
\mathcal{H}_{\mathrm{SU}(N)}
=t\Delta^{2}\mathcal{H}_{FG}-2N_p .
\end{equation}
Here,
\begin{equation}
\mathcal{H}_{FG}
=
\int \mathrm{d}x\,
\bigg[
(\partial_x\Psi_{\alpha}^{\dagger})
(\partial_x\Psi_{\alpha})
+
c\sum_{\alpha<\beta}^{N}
\Psi_{\alpha}^{\dagger}
\Psi_{\beta}^{\dagger}
\Psi_{\beta}
\Psi_{\alpha}
\bigg].
\end{equation}
The indices $\alpha$ and $\beta$ label the different SU($N$) components,
and $c=U/(t\Delta)$. The resulting continuum field theory corresponds
to the Gaudin--Yang--Sutherland model.

Its eigenstates may be expressed as
\[
\ket{\psi(\lambda)}
=
\sum_{\alpha_1,\ldots,\alpha_{N_p}=1}^{N}
\int
\chi(\mathbf{x}\mid\lambda)
\Psi_{\alpha_1}^{\dagger}(x_1)
\cdots
\Psi_{\alpha_{N_p}}^{\dagger}(x_{N_p})
\ket{0}\,\mathrm{d}\mathbf{x},
\]
where $\chi(\mathbf{x}\mid\lambda)$ is an eigenfunction of the
Gaudin--Yang--Sutherland Hamiltonian
\begin{equation}
\label{eq:hh4}
\mathcal{H}_{GYS}
=
-\sum_{\alpha=1}^{N}
\sum_{i=1}^{N_{\alpha}}
\frac{\partial^{2}}{\partial x_{i,\alpha}^{2}}
+
4c\sum_{i<j}\sum_{\alpha,\beta}
\delta(x_{i,\alpha}-x_{j,\beta}).
\end{equation}

The exact spectrum of the Gaudin-Yang-Sutherland model can be written as
$E=\sum\limits_{m=1}^{N}\sum\limits_{j=1}^{N_{m}}(mp_{j,m}^{2} - I_{m})$, following  from the string solutions of the Bethe equations  of the  model (\ref{eq:hh4}). Working in units of $\hbar^{2}/2m = 1$, the Bethe Ansatz energy is $E = \sum_{j}^{N_{p}}k_{j}^{2}$. For attractive interactions, in the limit of $L|c|\!\gg\!1$ with $L$ and $c$ being the length of the ring and interactions respectively, the charge rapidities organize themselves $k-\Lambda$ strings whilst the spin rapidities form $\Lambda$ strings. A bound state composed of $m$ particles is described by the following string configuration for the charge rapidities
\begin{equation}\label{eq:string}
    k_{j,\alpha}^{(m)}= p_{j,m}+\imath(m+1-2\alpha)c +\mathcal{O}(i\delta |c|),\qquad\alpha=1,\ldots,m ,
\end{equation}
where $p_{j,m}$ is the real part of the string in the nested Bethe Ansatz picture \cite{Takahashi1970,Sutherland1968,guan2013fermi}, $j=1,\ldots,N_{m}$ with $N_{m}$ being the number of $m$-body bound states. Naturally, we require that the occupation numbers satisfy $\sum_{m}^{N}mN_{m} = N_{p}$. The exponentially small deviations $\mathcal{O}(\imath\delta |c|)$, with $\delta\sim e^{-L|c|}$, are neglected in the ideal-string limit. \\

\noindent The energy contribution for a single bound state of size $m$ is given by
\begin{align}
    E_{j}^{(m)} &=\sum_{\alpha=1}^{m}\left[p_{j,m}+ \imath(m+1-2\alpha)c\right]^{2}  \\
    &=m p_{j,m}^{2}+2 \imath p_{j,m} c
    \sum_{\alpha=1}^{m}(m+1-2\alpha)-c^{2}\sum_{\alpha=1}^{m}(m+1-2\alpha)^{2}.
\end{align}
Seeing as the imaginary shifts are symmetric around zero as per \eqref{eq:string}, the mixed term in the above equation vanishes, $\sum_{\alpha=1}^{m}(m+1-2\alpha)=0 $ leaving us with 
\begin{equation}
    E_{j}^{(m)}=m p_{j,m}^{2}-I_{m},
    \qquad
    I_{m} =
    \frac{m(m^{2}-1)}{3}c^{2}.
\end{equation}
Lastly, if we sum over all the different sized bound states we get that
\begin{equation}
    E=\sum_{m=1}^{N}\sum_{j=1}^{N_m}\left(m p_{j,m}^{2}-I_m\right).
\end{equation}
From this expression, the hierarchy of the sub-bands can be directly related to the degree of clustering of the bound states. The lowest sub-band corresponds to
maximally bound $N$-body clusters, described by the longest $k-\Lambda$ strings, and therefore carries the largest negative binding contribution, $-I_{N}$. By contrast, higher sub-bands arise from the progressive breaking of these maximally bound clusters into lower-order bound states. Since shorter strings have smaller binding energies $I_{m}$, their negative contribution to the total energy is reduced, placing the corresponding states at higher energies.

\subsection{Characterizing bands through correlations}

\begin{figure}[H]
    \centering
{\includegraphics[width=0.8\columnwidth]{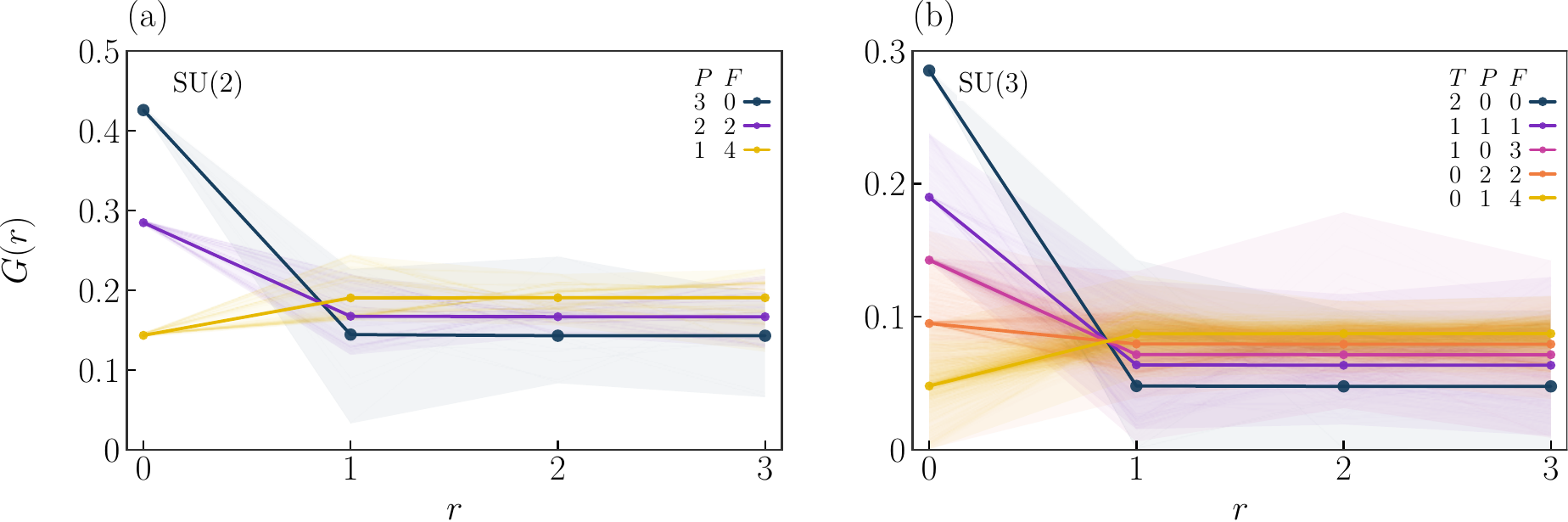}}
    \caption{Cross-flavor correlator $G(r)$ for individual eigenstates in each band, for all momentum sectors, for $L=7$, $|U|/J=20$. Bold lines are the band averages. For legibility only every second momentum sector is included, and within each band every tenth eigenstate is plotted. Curves are colored by the composite configuration of their band. The bound band has the sharpest $r=0$ peak and the tightest clustering, while higher bands are progressively flatter, justifying the band-averaged $G(r)$ across one momentum sector in Fig.~\ref{fig:fig1}.}
    \label{fig:collapse_appendix}
\end{figure}

For the SU($N$) Hubbard model presented in the main text, we probe the internal structure of each band with the cross-flavor correlator $G(r)$, which measures the probability that two different flavors are separated a distance $r$ apart. Figure~\ref{fig:collapse_appendix} shows the cross-flavor correlator $G(r)$ resolved for the individual eigenstates of each band, across all momentum sectors, at $L=7$ and $|U|/J = 20$. The curves are grouped up by the composite configuration of their band following the assignment of Fig.~\ref{fig:bandcomposition}. The bound band has the strongest peak at $r=0$, and the peak decreases from band to band as the number of flavors participating in bound clusters decreases.

The band count follows from the allowed partition of the particles into composites. For SU($2$) with $N_p = 6$ this gives four bands, ranging from three pairs to six unbound particles. For SU($4$) with $N_p = 4$ there are five bands, with the full bound quartet lying well below the SU($2$) and SU($3$) ground bands at the same $|U|/J$. For SU(3) with $N_p = 6$, the Bethe Ansatz partition predicts seven configurations, but the lattice spectrum resolves only six. The missing band arises because the configurations consisting of one trion plus three unbound particles and three pairs contain the same number of attractive flavor pairs and therefore have the same strong-coupling energy $\binom{m}{2}|U|$. In contrast, the continuum Bethe ansatz distinguishes these configurations through their different string structures: the trion and pair configurations have different binding energies, $-8c^2$ and $-6c^2$ respectively. Thus, these two configurations are degenerate at leading order in the lattice strong-coupling limit, where the Hubbard energy counts only the number of onsite attractive flavor pairs. This degeneracy need not be exact at finite $J/|U|$, but the resulting splitting can be small enough that the lattice spectrum resolves the two configurations as a single band. In contrast, the continuum Bethe ansatz distinguishes the corresponding string configurations already at the level of their binding energies.

\subsection{Composition of the bands}

\begin{figure}[H]
\centering
\includegraphics[width=.66\columnwidth]{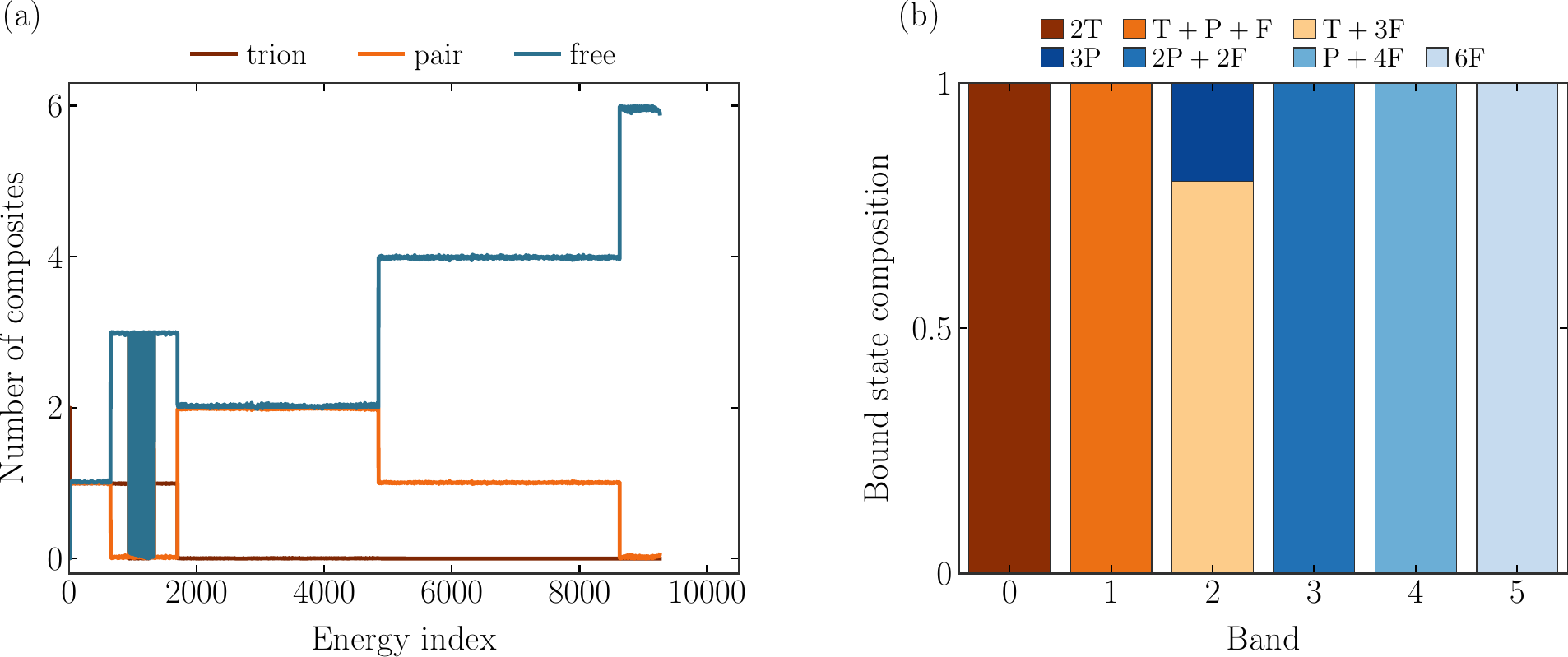}
    \caption{Composite structure of the SU($3$) many-body spectrum for $N_p = 6$ on $L = 7$ sites for interaction $|U|/J = 20$. \textbf{(a)} Number of trions $N_T$, pairs $N_P$, and free fermions $N_S$ as a function of eigenstate energy index (states ordered by increasing energy). Each count is flat within a band and jumps at the band edges as composites dissociate with increasing energy, from the fully bound band (two trions) to the full unbound highest band (six free fermions $N_S = 6$). \textbf{(b)} Configuration structure of each band, shown as the fraction of states with a given composite content ($T$ = trion, $P$ = pairs, $F$ = free fermion). Every band is a single configuration except band-2, which is a mixture of the $T+3F$ and $3P$ configurations, the degeneracy discussed in the text.}
    \label{fig:bandcomposition}
\end{figure}

These configurations can be identified directly by counting composites. At each site we measure whether the three flavors form a trion ($T_i$, all three present), a pair ($P_i$), or a (single) free fermion ($S_i$), and sum over the lattice to obtain the composite numbers $N_T = \sum_i \langle T_i \rangle$, $N_P = \sum_i \langle P_i \rangle$, $N_S = \sum_i \langle S_i \rangle$. Figure \ref{fig:bandcomposition} shows these across the spectrum: each stays flat within a band and jumps at the band edges, as the composite structure changes with increasing energy. Panel (b) shows the fraction of states band by band, where every band is a single configuration except the third, which mixes one trion, three singles and three pairs, as discussed above. 

\subsection{Gap $\Delta_j$}

\begin{figure}[H]
    \centering
{\includegraphics[width=0.56\columnwidth]{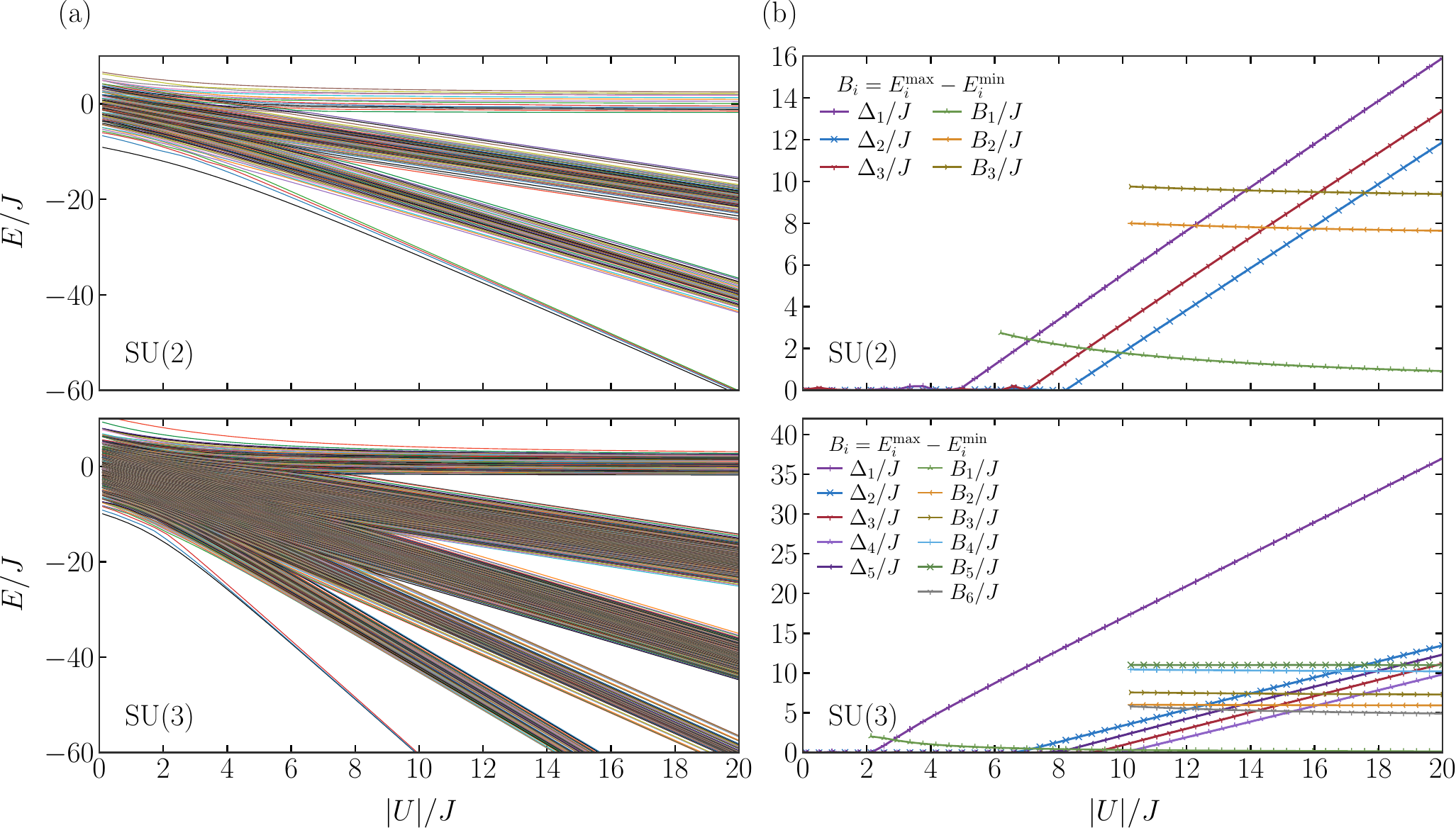}}
    \caption{Many-body energy spectrum $E(k)$ (left) and band gaps $\Delta_j/J$ and bandwidths $B_j/J$ (right) for SU(2) with $N_p=6$ (top) and SU(3) with $N_p=6$ (bottom), $L=7$. The gaps $\Delta_j$ separating consecutive bands open at a critical interaction strength and grow with $|U|/J$. For each SU($N$), the largest gap $\Delta_1$ separates the bound band from the first partially bound configuration. It opens first and grows fastest, consistent with the energy cost of dissociating the largest composite scaling linearly in $|U|$. Smaller gaps between higher partial bands open at larger $|U|/J$ and grow more slowly, consistent with the smaller energy cost of rearranging fewer bound fermions in the less-clustered region.}
    \label{fig:spectrum_gaps_appendix}
\end{figure}

Figure~\ref{fig:spectrum_gaps_appendix} shows the many-body energy spectrum and the interaction dependence of the band gaps and bandwidths. The gaps separating consecutive bands open at critical interaction strengths and subsequently increase with $|U|/J$.

\section{Dynamics}
\subsection{Effective hopping of the bound composite}
\label{sec:jtilde}

We perform a perturbative analysis of the composite dynamics as evidence that a maximally bound $N$-body composite propagates as a single composite with a re-normalized hopping amplitude $\tilde J$. The analysis applies specifically to one particle per component ($N_p/N = 1$), for which the derivation below holds. In the strongly attractive regime $|U|\gg J$, the bound composite is tightly localized in relative coordinate, with its constituents predominantly occupying the same lattice site. On a ring, however, its center of mass is not pinned to any particular site. The $L$ configurations in which the composite is centered on different lattice sites are energetically equivalent and combine into translationally invariant momentum eigenstates. The ground state corresponds to the zero-momentum superposition of the composite over all sites. Treating the hopping as a perturbation, the effective Hamiltonian to second order reads $\hat H_{\rm eff} = \hat P\hat V\hat P + \hat P\hat V\hat Q\,\frac{1}{E_0-\hat H_0}\,\hat Q\hat V\hat P+\dots,$ where $\hat P$ projects onto the degenerate manifold, $\hat Q=1-\hat P$ onto the split configurations outside it, and $\hat V=-J\sum_{i,\sigma}\left(\hat c^\dagger_{i+1,\sigma}\hat c_{i\sigma}+\mathrm{h.c.}\right)$ is the nearest-neighbor hopping, which moves a single fermion of flavor $\sigma$ between adjacent sites. 

\textit{Pairs ($N=2$).} The bound pair occupies a single site, $\ket{i}=\hat c^\dagger_{i\uparrow}\hat c^\dagger_{i\downarrow}\ket{0}$, with energy $E_0 = - |U|$. Its hopping to the neighboring site follows from the second-order matrix element $\bra{i+1}\hat H_{\rm eff}^{(2)}\ket{i}$ between the pair on site $i$ and on site $i+1$,
\begin{equation}
    \bra{i+1}\hat H_{\rm eff}^{(2)}\ket{i}
    = \sum_{m}\frac{\bra{i+1}V\ket{m}\bra{m}V\ket{i}}{E_0-E_m}.
\end{equation}

Two paths connect the pair on site $i$ to the pair on site $i+1$ since the $\uparrow$ and $\downarrow$ fermions are distinguishable. The pair can split with either fermion hopping first, $\ket{m_A}=\ket{\uparrow_{i+1}\,\downarrow_i}$ or $\ket{m_B}=\ket{\downarrow_{i+1}\,\uparrow_i}$. Both are configurations with the two fermions on different sites, so they carry no on-site pair and $E_m=0$. Each hop is generated by $V=-J\sum_{i\sigma}(\hat c^\dagger_{i+1\sigma}\hat c_{i\sigma}+\mathrm{h.c.})$ and contributes an amplitude $-J$. For each path the second hop carries a fermionic sign from moving one operator past the other, but these cancel between the two hops, leaving a net $+1$, so the paths add in phase. Each therefore contributes
\begin{equation}
    \frac{(-J)(-J)}{E_0-E_m}=\frac{J^2}{-|U|}=-\frac{J^2}{|U|}.
\end{equation}

Summing the two paths gives the effective pair hopping,
\begin{equation}
    \bra{i+1}\hat H_{\rm eff}^{(2)}\ket{i}
    = -\frac{2J^2}{|U|}\equiv-\tilde J,
    \qquad
    \tilde J=\frac{2J^2}{|U|},
\end{equation}
where the factor of $2$ counts the two equivalent hopping paths.

\textit{Trions ($N=3$).} The bound trion occupies a single site, $\ket{i}=\hat c^\dagger_{ia}\hat c^\dagger_{ib}\hat c^\dagger_{ic}\ket{0}$, with energy $E_0=-3|U|$, and now translates at third order, 
\begin{equation}
    \bra{i+1}\hat H_{\rm eff}^{(3)}\ket{i}
    = \sum_{\rm paths}
    \frac{\bra{i+1}\hat V\ket{m_2}\bra{m_2}\hat V\ket{m_1}\bra{m_1}\hat V\ket{i}}
         {(E_0-E_{m_1})(E_0-E_{m_2})},
\end{equation}
through the two intermediate states. As before the flavors hop sequentially and the fermionic signs cancel, so all paths add in phase. Each intermediate contains one onsite pairwise attractive interaction ($E_{m} = -|U|$) giving denominators $E_0-E_m=-2|U|$. The three distinguishable flavors may cross in $3! = 6$ orders, so 
\begin{equation}
    \bra{i+1}\hat H_{\rm eff}^{(3)}\ket{i}
    = 6\,\frac{(-J)^3}{(-2|U|)^2}
    = -\frac{3}{2}\frac{J^3}{|U|^2}\equiv-\tilde J,
    \qquad
    \tilde J=\frac{3}{2}\frac{J^3}{|U|^2}.
\end{equation}

\noindent Both cases follow the general form 
\begin{equation}
\label{eq:jtilde_general}
    \tilde J \simeq J\,\frac{N}{(N-1)!}\left(\frac{J}{|U|}\right)^{N-1}.
\end{equation}
Thus, we note that  for fermions with $N_p/N=1$ where $N$ is the number of components,  the  bound  particles move with  an effective hopping scaling $(J/|U|)^{N-1}$, similarly to the bosonic case~\cite{mansikkamaki2022beyond,blain2026quantum}.

\subsection{Expansion velocity $v_\infty$}

\begin{figure}[H]
    \centering
    \includegraphics[width=0.5\columnwidth]{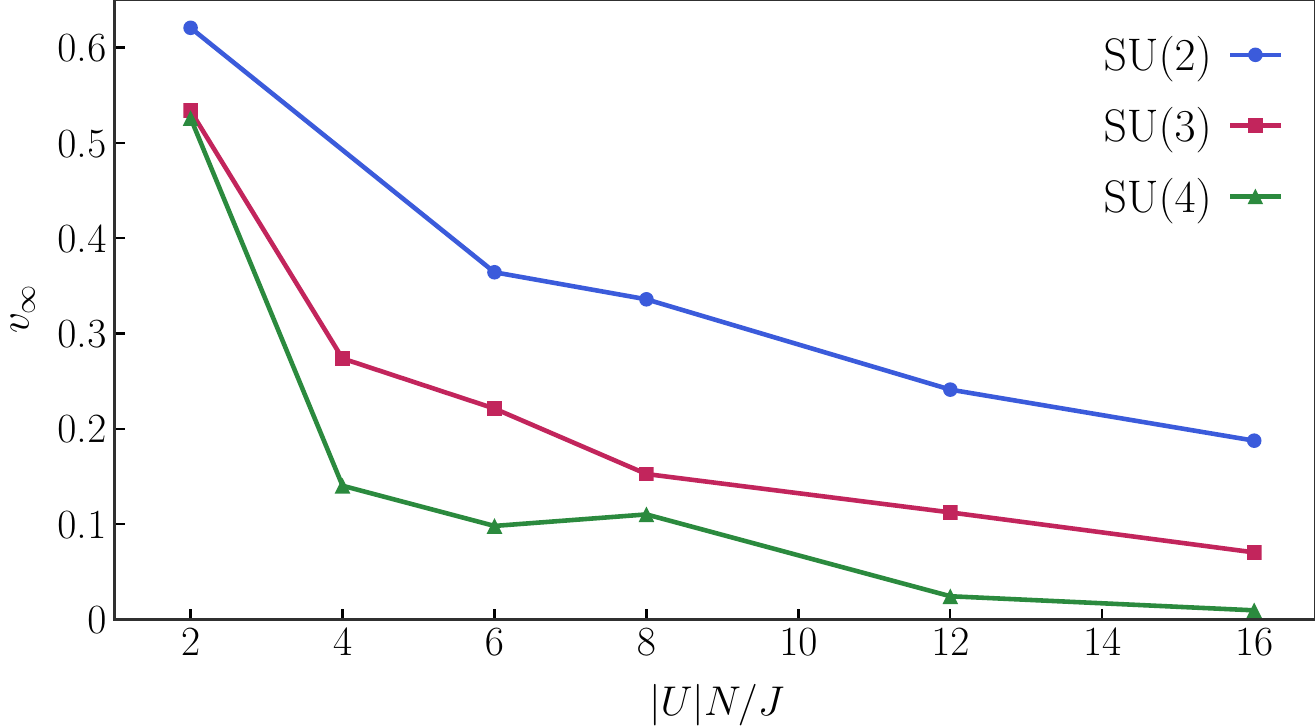}
    \caption{Expansion velocity $v_\infty$ versus interaction strength $|U|N/J$ with $N_p/N = 1$ on $L=9$ sites. $v_\infty$ is the
asymptotic spreading velocity extracted from $\sqrt{R^2(t)-R^2(0)}$, averaged
over the window preceding the finite-size wraparound cutoff $t^*$. The velocity
is suppressed with increasing $|U|N/J$, and the suppression strengthens with $N$.}
    \label{fig:velocity}
\end{figure}

Figure~\ref{fig:velocity} shows the asymptotic expansion velocity $v_\infty$ of the bound composite as a function of the interaction strength $|U|N/J$, for $N_p/N = 1$. We extract $v_\infty$ from the spreading of the density profile. We first measure the mean squared displacement of the density about the pinning site $j_0=(L-1)/2$,
\begin{equation}
  R^2(t) = \frac{1}{N}\sum_{j=1}^{L} n_j(t)\,(j - j_0)^2,
\end{equation}
which grows as the composite expands across the lattice. From its square root we obtain the spreading width and its corresponding velocity. 
\begin{equation}
    \sigma(t) = \sqrt{R^2(t) - R^2(0)}, \qquad v(t) = \frac{d\sigma}{dt}
\end{equation}
We then report the expansion velocity $v_\infty$ as the average of $v(t)$ over the late ballistic window to capture the steady expansion rate, before the wavefront wraps around the finite ring. We plot against $|U|N/J$ rather than the bare interaction $|U|$ because it matches the binding strength across different $N$: at fixed $|U|N/J$ the composites of different flavor number are compared equally, so the differences between the curves reflect the physics rather than the difference in binding. For all three SU($N$), $v_\infty$ decreases with increasing $|U|N/J$: stronger attraction binds the composite more tightly, increasing the effective mass and slowing its spread across the lattice. At fixed $|U|N/J$, the velocity decreases with components N, SU($4$) being the slowest. This ordering follows from the effective hopping $\tilde J \simeq J\,\frac{N}{(N-1)!}\left(\frac{J}{|U|}\right)^{N-1}$: larger $N$ gives a smaller $\tilde J$, hence a less mobile composite that is also heavier. That the composite moves intact, traveling as one object rather than breaking apart, is shown by the density evolution and Hellinger analysis in Fig.~\ref{fig:fig2}.

\subsection{Disorder robustness of the $N$-body correlations}

\begin{figure}[H]
    \centering
\includegraphics[width=\columnwidth]{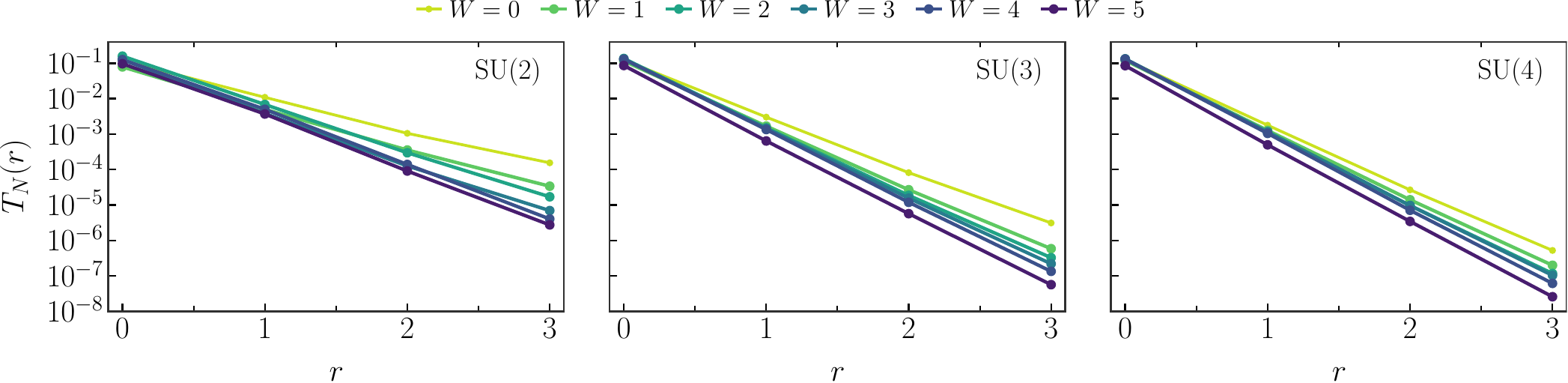}
    \caption{$N$-body correlator $T_N(r) = \langle \hat n_{j_0+r,1}\prod_{\alpha=2}^{N}
    \hat n_{j_0,\alpha}\rangle$ (log scale) versus distance $r$ for the disorder ground state, for SU(2), SU(3), and SU(4), at one particle per component ($N_p/N = 1$) and matched interaction $|U|N/J$ = 12, on $L = 7$ sites. Within each panel the curves correspond to increasing disorder strength $W/J = 0,1,\dots,5$ (light to dark), each averaged over disorder realizations. $T_N(r)$ decays exponentially and the decay generally steepens with $W$ (the correlation length $\xi_N$ shrinks); the steepening is stronger for larger $N$, consistent with the decay rates $1/\xi_N$ in Fig.~\ref{fig:fig3}.}
    \label{fig:nbody_disorder}
\end{figure}

The inverse correlation lengths $1/\xi_N$ in Fig.~\ref{fig:fig3} are extracted from exponential fits to the disorder-averaged ground-state $N$-body correlators $T_N(r)$. Figure~\ref{fig:nbody_disorder} shows these correlators directly. For every SU($N$), $T_N(r)$ decays exponentially in $r$, and more steeply as $W$ grows, so the $N$-body correlations become more tightly confined in relative distance under stronger disorder. At fixed $W$, the fall-off is steeper for larger $N$; higher-$N$
composites are more strongly confined by disorder, as summarized by $1/\xi_N$ in Fig.~\ref{fig:fig3}.

\section{Dynamics in the general case of $N_p/N > 1$}
\begin{figure}[H]
    \centering
    \includegraphics[width=0.9\columnwidth]{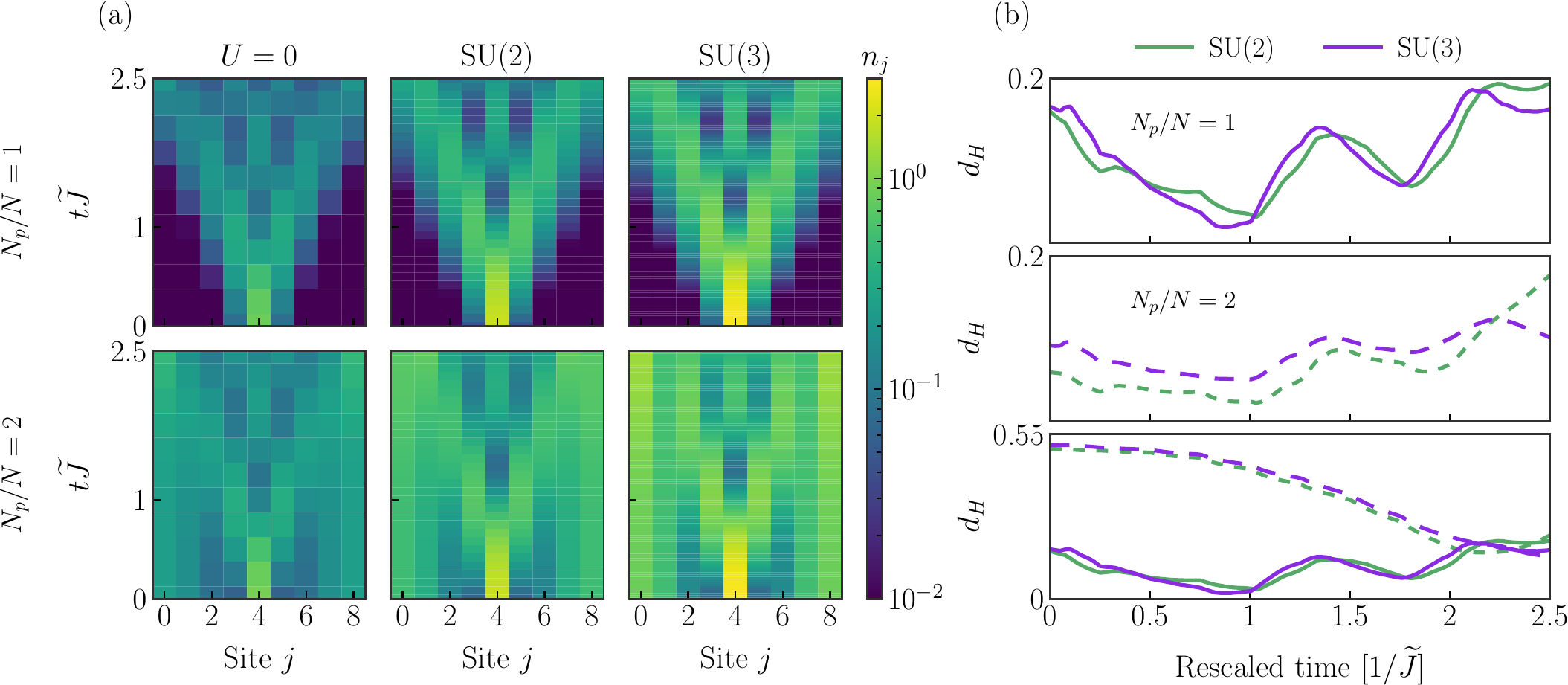}
    \caption{Comparison of $N_p/N=1$ and $N_p/N=2$ on $L=9$ sites at matched interaction $|U|N/J = 12$. \textbf{(a)} Density evolution $n_j$ (log scale) in rescaled time $\tau = t\tilde J$ after release of pinning potential, for the non-interacting case ($U = 0$), SU(2) and SU(3) (columns), with $N_p/N = 1$ (top row) and $N_p/N = 2$ (bottom row). \textbf{(b)} Hellinger distance $d_H$ between the evolving density and the free-fermion reference as a function of the rescaled time. Solid lines are $N_p/N = 1$ [SU(2) green, SU(3) purple], the $N_p/N = 2$ curves are dotted for SU(2) and dashed for SU(3). The $N_p/N = 2$ distance is larger than the $N_p/N = 1$ distance, and the time where they cross marks the onset of self-interference after the wave packet travels around the two arms of the ring.}
    \label{fig:evolution_npn2}
\end{figure}

Here, we compare the expansion dynamics after pinning states with $N_p/N=1$ and $N_p/N=2$ - Fig.\ref{fig:evolution_npn2}. 
For the states  $N_p/N=1$, all the particles  can be pinned in  a single site.  
For states with  $N_p/N=2$ instead, Fig.\ref{fig:evolution_npn2} indicates that the initial states contain both bound and un-bound particles. 

  In particular, we note that the  dynamics corresponding to $N_p/N=2$ is generically characterized by a larger  Hellinger distance compared with the $N_p/N=1$. Such feature provides a characterization of the $N_p/N=1$ states as quantum walks. We comment that the  time at which the two distances cross marks the onset of self-interference of the wave packet after traversing the two  circular arms of the ring. 
This behavior is indicative of suppressed self-interference in the solitary wave compared with the interference exhibited by the unpaired particles present in the $N_p/N=2$ states.

\label{sec:appendix}

\end{document}